\documentclass{aa}
\usepackage{graphicx,amssymb}

\def\mes{M{\'e}sz{\'a}ros}

\begin{document}

\title{Polarimetric observations of GRB~011211\thanks{Based on observations 
made with ESO telescopes at the Paranal Observatories under programme Id
68.D-0064.}}
\titlerunning{Polarization of GRB~011211}

\author{S. Covino\inst{1}, D. Lazzati\inst{2}, D. Malesani\inst{1},
G. Ghisellini\inst{1}, 
G. L. Israel\inst{3}, L. Stella\inst{3}, A. Cimatti\inst{4}, S. di
Serego\inst{4}, F. Fiore\inst{3}, N. Kawai\inst{5},
S. Ortolani\inst{6}, L. Pasquini\inst{7}, G. Ricker\inst{8},
P. Saracco\inst{1}, G. Tagliaferri\inst{1}, F. Zerbi\inst{1}}

\authorrunning{Covino et al.}

\offprints{S. Covino}
\institute{Osservatorio Astronomico di Brera, via E. Bianchi 46, 
23807 Merate (LC), Italy
\and
Institute of Astronomy, University of Cambridge, Madingley
Road, CB3 0HA Cambridge, England 
\and
Osservatorio Astronomico di Roma, via Frascati 33, Monteporzio
Catone (Roma), Italy 
\and
Osservatorio Astrofisico di Arcetri, Largo E. Fermi 5, 50125 Firenze,
Italy
\and 
Dept. of Physics, Tokyo Institute of Technology,
2-12-1 Ookayama, Meguroku, Tokyo 152-8551, Japan
\and 
Universit\`a di Padova, Dipartimento di Astronomia, Vicolo
dell'Osservatorio 2, 35122 Padova, Italy
\and 
European Southern Observatory, Karl Schwarzschild Strasse 2, 85748
Garching bei M\"unchen, Germany
\and
Center for Space Research, Massachusetts Institute of Technology,
Cambridge, Massachusetts 02139--4307, USA}

\date{}

\abstract{
We present and discuss polarimetric observations performed with the
VLT--UT3 (Melipal) on the afterglow of GRB~011211, $\sim35$~hours
after the burst onset. The observations yielded a 3-$\sigma$ upper
limit of $P<2.7\%$. We discuss this result in combination with the
lightcurve evolution, that may show a break approximately at the time
of our observation. We show that our upper limit is consistent with
the currently favored beamed fireball geometry, especially if the line
of sight was not too close to the edge of the cone.
}

\maketitle

\section{Introduction}
\label{sec:int}

It is now generally believed that the afterglow ubiquitously observed
in GRBs is produced by synchrotron radiation (see, e.g., Piran \cite{Pi99})
as a beamed relativistic fireball is decelerated by the impact with
the ambient medium ({\mes} \& Rees \cite{MR97}). This interpretation is
confirmed by the observation of power-law decaying lightcurves (Wijers
et al. \cite{WRM97}) showing a break at $t \sim 1$-30~days (Frail et
al. \cite{Fr01}), of power-law spectral energy distributions (Wijers \&
Galama \cite{WG99}; Panaitescu \& Kumar \cite{PK01}) and of linear polarization
(Covino et al. \cite{Co99}; Wijers et al. \cite{Wi99}; Rol et al. \cite{Rol00}).

The derivation of the fireball opening angle from the time of breaks
in the afterglow lightcurves is crucial to derive the energy
budget of GRBs (Frail et al. \cite{Fr01}). It is nevertheless a matter of
open debate whether the breaks are due to collimation or to different
hydrodynamical transitions (Moderski et al. \cite{Mo00}; in 't Zand et
al. \cite{iZ01}). The presence of polarization, and in particular its
evolution (Ghisellini \& Lazzati \cite{GL99}, hereafter GL99; Sari \cite{Sari99})
is an alternative and unbiased way to prove that the fireball is beamed
and allows to constrain the orientation of the jet with respect to the
line of sight to the observer (GL99; Bj\"ornsson \& Lindfors \cite{BL00}).

Before the observation presented here, 4 GRBs have been observed in
polarimetric mode at various wavelengths, yielding two positive
measurements and two upper limits.  The first measurement was
performed on the afterglow of GRB~990123 in the $R$ band, yielding an
upper limit $P < 2.3\%$ ($95\%$ confidence level, Hjorth et al. \cite{Hj99}).
The first detection of linear polarization was obtained by Covino et
al.  (\cite{Co99}) on GRB~990510. Observations in the $R$ band at
$t\sim18.5$~hours after the burst yielded $P = (1.7\pm0.2)\%$. The
detection was confirmed by Wijers et al. (\cite{Wi99}), who obtained
$P = (1.6\pm0.2)\%$ at $t \sim 20$~hours, a value consistent with that of
Covino et al. (\cite{Co99}). Multiple measurements of polarization at three
different epochs were performed on GRB~990712 (Rol et
al. \cite{Rol00}). While the position angle did not vary significantly (but the data 
are also consistent with a 45$^\circ$ variation), a marginal detection
of fluctuation of the polarized fraction was obtained, the second
measurement ($P = (1.2\pm0.4)\%$ at $t \sim 16.7$~hours) being smaller
than the other two ($P = (2.9\pm0.4)\%$ and $P = (2.2\pm0.7)\%$ at
$t \sim 10.6$~hours and $t \sim 34.7$~hours, respectively).  Finally,
an attempt to measure near infrared (NIR) polarization in the afterglow
of GRB~000301C yielded only a weak $P < 30\%$
constraint\footnote{Several other attempts to measure linear
polarization of afterglows in the NIR were performed by the same
collaboration, but it turned out that for all these bursts an
optical--IR afterglow was not detected.}  (Stecklum et al. \cite{St01}).

As a general rule, some degree of asymmetry is necessary in order to
observe polarization. Two general models have been proposed
to explain some degree of linear polarization in the
framework of synchrotron emission. Gruzinov \& Waxman (\cite{GW99}) discuss
how ordered magnetic field domains can diffuse in the fireball,
predicting $ P\sim 10\%$. GL99 (and, independently, Sari \cite{Sari99})
considered a geometrical setup in which a beamed fireball observed
slightly off-axis provides the necessary degree of anisotropy (see
also Sect.~\ref{sec:model}). Variable polarization up to $10\%$ is
predicted.

\section{Data and analysis}
\label{sec:data}

GRB~011211 was detected by {\it Beppo}SAX on Dec. 11, 19:09:21 UT and initially
classified as part of the X--ray rich class (Gandolfi \cite{Ga01}).
Refined analysis (Frontera et al. \cite{Fr02}) showed that it was
actually a standard GRB. The optical afterglow was discovered after 10~hours
(Grav et al. \cite{Gr01}) and confirmed by Bloom \& Berger (\cite{BB01}).

\begin{figure}
\centerline{\resizebox{\hsize}{!}{\includegraphics{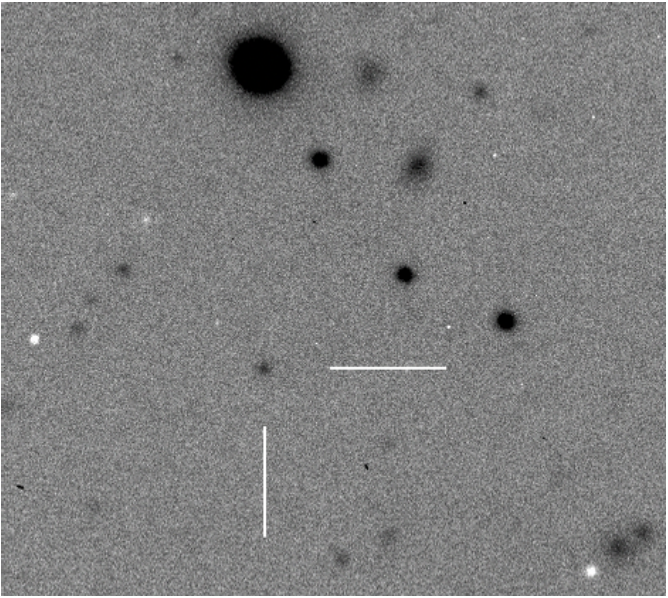}}}
\caption{The optical afterglow to GRB~011211 in the Bessel $V$-band
VLT--UT3 acquisition frame.}
\label{fig:OT}
\end{figure}

Our observations of GRB~011211 were obtained at ESO's VLT--UT3
(Melipal), equipped with the Focal Reducer/low dispersion Spectrometer
(FORS) and Bessel filter $R$. The OT associated with GRB~011211 was
observed for $\sim 3$~hours, starting $\sim 35$~hours after the burst,
when the $V$- and $R$-band magnitudes were $21.70 \pm 0.06$
and $21.43\pm0.1$, with
respect to the USNO star U0675\_11427359 (Covino et al. \cite{Co02}, Henden \cite{He02}). 
Observations were performed in standard resolution mode with a scale of $0.2\arcsec$/pixel 
(Fig.~\ref{fig:OT}); the seeing varied from $\sim 1.4\arcsec$ at the beginning to $\sim 
0.7\arcsec$ at the end. The observation log is reported in Table~\ref{tab:log}.

\begin{table}
\begin{center}
\begin{tabular}{ccccc}
\hline
\bf Starting time  & \bf Exposure & \bf Angle   & \bf Filter & \bf Seeing \\
\sf UT, 13 Dec 2001 & \sf sec     & \sf deg     &            & \sf arcsec \\
\hline
      05:40        &   720        &     00.0    & $R$        & 1.4        \\
      05:53        &   720        &     22.5    & $R$        & 1.2        \\
      06:06        &   720        &     45.0    & $R$        & 1.0        \\
      06:19        &   720        &     67.5    & $R$        & 1.0        \\
      06:34        &   720        &     00.0    & $R$        & 1.0        \\
      06:47        &   720        &     22.5    & $R$        & 0.9        \\
      07:00        &   720        &     45.0    & $R$        & 0.8        \\
      07:13        &   720        &     67.5    & $R$        & 0.8        \\
      07:45        &   720        &     00.0    & $R$        & 0.8        \\
      07:58        &   720        &     22.5    & $R$        & 0.8        \\
      08:15        &   720        &     45.0    & $R$        & 0.8        \\
      08:28        &   720        &     67.5    & $R$        & 0.7        \\
\hline
\end{tabular}
\end{center}
\caption{Observation log for the polarimetric observation of the 
GRB~011211 field.}
\label{tab:log}
\end{table}

Imaging polarimetry is achieved by the use of a Wollaston prism
splitting the image of each object in the field into the two
orthogonal polarization components which appear in adjacent areas of
the CCD image. For each position angle $\phi/2$ of the half--wave
plate rotator, we obtain two simultaneous images of
cross-polarization, at angles $\phi$ and $\phi + 90^\circ$.

Relative photometry with respect to all the stars in the field was
performed and each couple of simultaneous measurements at orthogonal
angles was used to compute the $U$ and $Q$ Stokes parameters. This technique
removes any difference between the two optical paths (ordinary and
extraordinary rays) and the polarization component introduced by
Galactic interstellar grains along the line of sight. Moreover, since 
the Stokes parameters are directly derived from the source intensity 
ratio between the ordinary and extraordinary beams which are recorded 
simultaneously, they are not influenced by intensity variations of the 
source, provided that the polarization remained constant during the exposure
time. If the polarization has varied, what is obtained is the average 
of the Stokes parameters during the measurement (for further details 
on the reduction algorithm applied to data obtained with a dual--beam 
instruments like the FORS1 see e.g. Cohen et al. \cite{CVO97} and di 
Serego Alighieri \cite{dS97}). 

With the same procedure, we observed also one polarimetric
standard star, Vela1\,95, in order to fix the offset between the
polarization and the instrumental angles.

The data reduction was carried out with the ESO-MIDAS (version 01SEP)
system. After bias subtraction, non--uniformities were corrected using
flat-fields obtained with the Wollaston prism. The flux of each point
source in the field of view was derived by means of both aperture and
profile fitting photometry by the DAOPHOT~II package (Stetson \cite{St87}),
as implemented in MIDAS. For relatively isolated stars the two
techniques differ only by a few parts in a thousand.

\begin{figure}
\centerline{\resizebox{0.8\hsize}{!}{\includegraphics{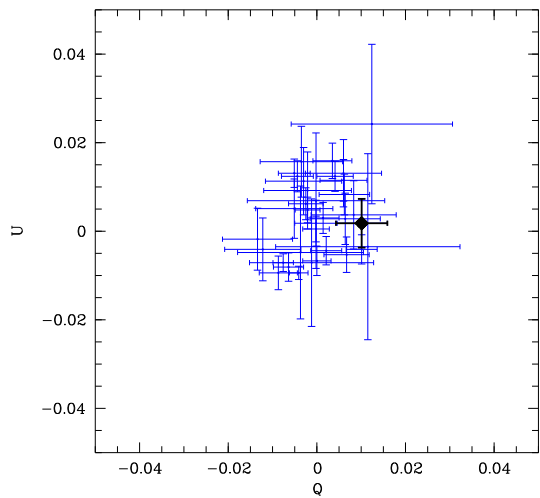}}}
\caption{Polarization normalized Stokes parameters $U$ and $Q$ for 
GRB~011211 optical transient (bold cross) and stars in the field.}
\label{fig:pol}
\end{figure}

In Fig.~\ref{fig:pol} we plot on the plane defined by the normalized
Stokes parameters $Q$ and $U$ the results of the polarization
measurements performed for the optical transient and for most of the
stars in the field of view. The average polarization of the stars
is consistent with zero: $\langle Q \rangle = -0.0015 \pm 0.0008$ and
$\langle U \rangle = -0.0007 \pm 0.0007$.  The normalized polarization
Stokes parameters for the optical transient are $Q = 0.0101 \pm 0.0058$ and
$U = 0.0018 \pm 0.0055$. The formal degree of polarization could in
principle be obtained from the measurements of $Q$ and $U$
($P = \sqrt {U^2 + Q^2}$) after correcting for the instrumental or local
interstellar polarization ($\langle Q \rangle$ and $\langle U \rangle$).
However, for very low
level of polarization ($P/\sigma \le 4$), a correction which takes
into account the bias due to the fact that $P$ is a definite
positive quantity (Wardle \& Kronberg \cite{WK74}) is required. At low
polarization level, the distribution function of $P$ (and of $\theta$,
the polarization angle) are no longer normal and that of $P$ becomes
skewed (Clarke et al. \cite{Cl83}, Simmons \& Stewart \cite{SS85}, Fosbury et
al. \cite{Fo93}). We therefore corrected the bias following Simmons \&
Stewart (\cite{SS85}) and derived a 3-$\sigma$ upper limit of $2.7\%$ ($2.0\%$
at $95\%$ confidence level) for the polarization degree of the optical
transient of GRB~011211. Monte Carlo simulations confirmed the
reported upper limits\footnote{In Covino et al. (\cite{Co02}) we reported a
preliminary 3-$\sigma$ upper limit slightly lower: $2.5\%$.}.

\section{Modeling}
\label{sec:model}

\begin{figure}
\centerline{\resizebox{\hsize}{!}{\includegraphics{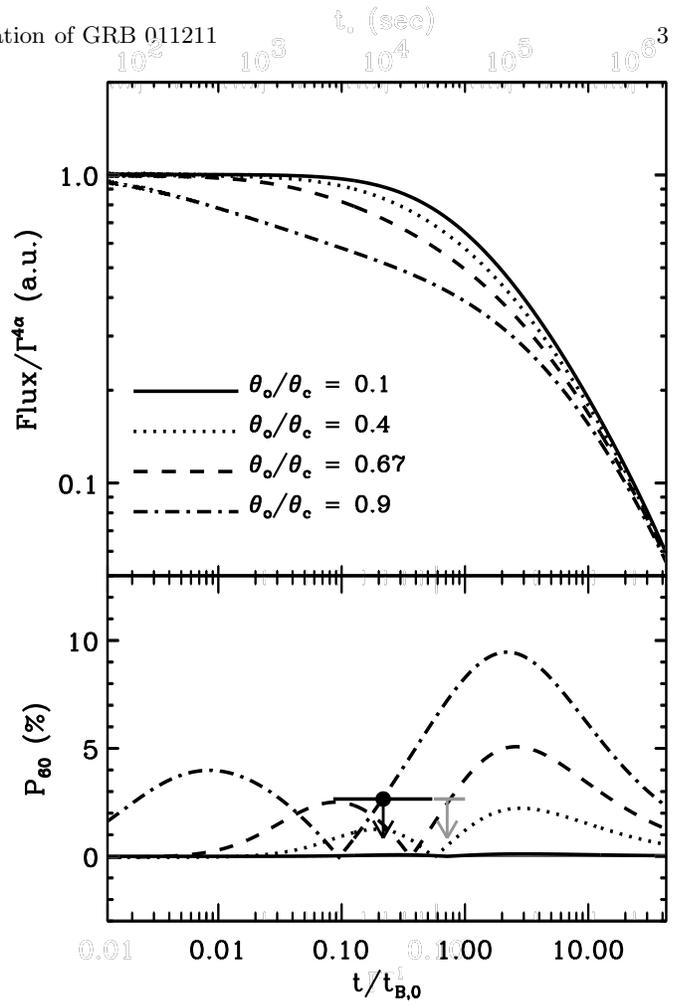}}}
\caption{{Lightcurves (upper panel) and linear polarization (lower panel)
of the OT as a function of the ratio of the observation time over the
break time $t_{\rm{B,0}}$, as measured by an observer along the
symmetry axis of the jet ($\theta_o/\theta_c = 0$). Different lines are
relative to different off-axis position of the line of sight (see text
and indications in the figure). The black upper limit shows the
position of our measurement if GRB~011211 follows the correlation of
Frail et al. (\cite{Fr01}), while the gray upper limit refers to the
lightcurve break detected by Holland et al. (\cite{Ho02}).}
\label{fig:tpol}}
\end{figure}

Linear polarization measurements have been performed, to date, in 5
afterglows. Even though theoretical models predict that the degree of
the polarization can be as high as $10\%$ (Gruzinov \& Waxman \cite{GW99};
GL99; Sari \cite{Sari99}), the afterglows seem to show only a few per cent of
polarization, if any.

In the model of Gruzinov \& Waxman (\cite{GW99}), a smaller polarization can
be explained by increasing the number of ordered magnetic field
domains $N_{\rm dom}$, since $P\sim60/\sqrt{N_{\rm dom}}$. In the
beamed fireball model, the polarization is due to the geometric
asymmetry provided by a beamed fireball observed off-axis. The degree
of polarization depends on the ratio of the angle between the line of
sight and the cone axis ($\theta_o$) to the opening angle of the jet
($\theta_c$). In addition, the degree of linear polarization is time
dependent, with two separate peaks (the first always smaller than the
second) spaced by a moment of null polarization. In this moment the
position angle of the polarization vector abruptly changes by
$90^\circ$. The expected degree of polarization can then be computed
by constraining the fireball geometry. In Fig.~\ref{fig:tpol} we show
the predictions of the model as a function of the ratio
$\theta_o/\theta_c$ and of the ratio $t/t_{\rm{B,0}}$, where $t$ is
the observed time and $t_{\rm{B,0}}$ is the break time that an
observer at $\theta_o = 0$ would measure in the lightcurve. Note that in
the original Fig.~4 of GL99, the polarization was shown as a function
of the inverse of the Lorentz factor. Since, however, both the break
time and the linear polarization are functions of the geometrical
properties of the jet only, the observed polarization is a function of
$t/t_{\rm{B,0}}$, without loss of generality (Sari \cite{Sari99}).

Holland et al. (\cite{Ho02}) claim the detection of a break in the
optical lightcurve of GRB~011211 at $1.5 < t < 2.7$~days; this is
confirmed also by later measurements at $t \gtrsim 10$~d (Burud et
al. \cite{Bu01}; Fox et al. \cite{Fox02}). If such break is indeed due to
collimation in the outflow, our polarimetric observation was performed at
$0.5 < t/t_{\rm{B,0}} \lesssim 1$. The upper limit is shown with a grey arrow
in Fig.~\ref{fig:tpol}. Our upper limit is then consistent with the model
prediction for $\theta_o < 0.67~\theta_c$. Since half of the random
oriented observers satisfy this constraint, our upper limit is fully
consistent with the theory of jetted fireballs.

\begin{figure}
\centerline{\resizebox{\hsize}{!}{\includegraphics{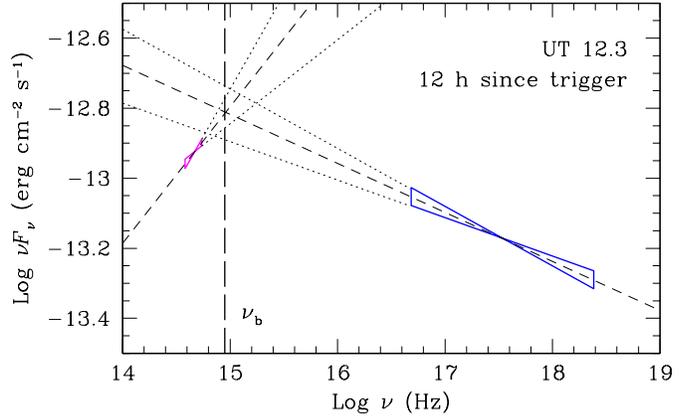}}}
\caption{{Broad-band optical to X--ray spectrum on Dec.~12.3 UT.
Optical data (corrected for Galactic extinction) are from Holland et
al. (\cite{Ho02}), while X-ray data are from Borozdin \& Trudolyubov
(\cite{BT02}). The low- and high-energy spectral index ($F_\nu \propto
\nu^{-\alpha}$) are $\alpha_{\rm o} = 0.6 \pm 0.15$ and $\alpha_{\rm X} =
1.1 \pm 0.03$ (note that the plot is in $\nu F_\nu$, so the slopes are
diminished by one unity)}. A spectral break is present at
$\nu_\mathrm{b} \sim 10^{15}$~Hz.
\label{fig:spec}}
\end{figure}

However, the analysis of the broad-band spectrum taken on Dec.~12.3
($\sim 1$ day before our polarization measurement),  
including data in the optical (Holland et al. \cite{Ho02}) and
X--ray bands (Reeves et al. \cite{Re02}; Borozdin \& Trudolyubov
\cite{BT02}), requires the presence of a spectral break at about
$\nu_\mathrm{b} \sim 10^{15}$~Hz, very close to the optical band
(Fig.~\ref{fig:spec}).
In the context of the standard synchrotron model (Sari, Piran \& Narayan,
\cite{SNP98}), this can be interpreted either as the injection frequency
(in the fast cooling regime) or the cooling frequency (in the slow cooling
regime). In the first case, the low-energy spectral index should be
$\alpha_{\rm o} = 0.5$, consistent with the observed value $0.6 \pm 0.15$,
while in the second case the difference between the high- and low-energy
slopes should be $\alpha_{\rm X} - \alpha_{\rm o} = 0.5$, also consistent
with the observed one $0.53 \pm 0.15$. Most afterglow models predict that
$\nu_\mathrm{b}$ should decrease with time, yielding a {\it chromatic} break
in the lightcurve, expected soon after Dec.~12.3. The time needed for
$\nu_{\rm b}$ to pass through the optical band is $\sim 1$~d, and the
lightcurve is not sampled enough to discriminate between the chromatic and
achromatic case. 
The case for a chromatic break receives some support from the analysis
of the temporal behaviour of the afterglow. In fact, the observed
flux decreases with time, following the power-law trend
$F_\nu \propto t^{-\delta_\nu}$, where $\delta_{\rm o} = 0.83 \pm 0.04$ in
the optical band and $\delta_{\rm X} = 1.6 \pm 0.1$ in the X--ray band
(Holland et al. \cite{Ho02}; Borozdin \& Trudolyubov \cite{BT02}). The
change in the decay slope after the passage of $\nu_{\rm b}$ through the
optical band is hence predicted to be $\Delta = \delta_{\rm X} -
\delta_{\rm o} = 0.77 \pm 0.1$, fully consistent with the value observed
by Holland et al. (\cite{Ho02}): $\Delta_{\rm obs} \ge 0.6$.

We can derive a second estimate for the jet break time using the
energy vs break time correlation (Frail et al. \cite{Fr01}). 
Using Fig.~3 of Bloom, Frail \& Sari (\cite{BFS01}), we derive a 
bolometric isotropic energy $E_{\rm{iso,bol}} \sim 10^{53}$~erg from 
the (40-700)~keV energy release $E_{\rm{iso}} \sim
6\times10^{52}$~erg (Frontera et al. \cite{Fr02}).
We can then estimate the expected jet--break time, which
turns out to be (allowing for a factor of two uncertainty in the
beaming-corrected total energy), $3 < t_{\rm B} < 18$~d.
This time is therefore  much later than the time $t_{\rm pol}\sim1.5$~d
at which the polarization measurement was performed.
This estimate of jet-break time, converted into $t/t_{\rm{B,0}}$, is
shown by the black arrow in Fig.~\ref{fig:tpol}. This figure shows that
the polarization measurement was probably  performed when the polarized
fraction was at its minimum for possibly all the fireball configurations.

\section{Conclusions}
\label{sec:con}
We have observed in polarimetric mode the optical afterglow of
GRB~011211; our result is a 3-$\sigma$ upper limit of $P <
2.7\%$. This is consistent with previous measurements performed on
other GRBs. Unfortunately a clear achromatic jet break is not observed
in the burst lightcurve, and this does not allow us to perform a clear
comparison with the currently favored theoretical models for the
production of polarization in beamed fireballs. We can nevertheless
deduce that, if the ratio of the observing angle to the jet opening
angle was less than 2/3, our measurement would be consistent with the
models. This result holds true if a break was present at $t\sim2$~days
(Holland et al. \cite{Ho02}) or if it was at a much later time.

\begin{acknowledgements}
We thank the anonymous referee for her/his prompt reply and detailed reading of our manuscript.
\end{acknowledgements}

\end{document}